\documentstyle[12pt]{article}

\begin{document}

\hsize 15.5 cm

\centerline{\Large Kolmogorov Dispersion for Turbulence in Porous Media:}

\centerline{\Large A Conjecture}

\bigskip

\centerline {Bikas K. Chakrabarti}
\centerline {Saha Institute of Nuclear Physics}
\centerline {1/AF Bidhan Nagar, Kolkata-700064, India.}

\bigskip

\bigskip

\noindent {\small {\bf Abstract:}
  We will utilise the self-avoiding walk (SAW) mapping
of the vortex line conformations in turbulence to get the Kolmogorov
scale dependence of energy dispersion from SAW statistics, 
and the knowledge of the disordered
fractal geometries on the SAW statistics. These will give us the 
Kolmogorov energy dispersion exponent value for turbulence in porous media 
in terms of the size exponent for polymers in the same. We
 argue that the exponent value will be somewhat less than 5/3
for turbulence in porous media.}

\vskip 1 cm

\noindent The turbulent flow of a fluid is a well known  stochastic 
process with dispersion, where the energy transfers occur in various 
modes corresponding to different length scales. Based on the general properties
of the Navier-Stokes equation of hydrodynamics and dimensional
analysis, Kolmogorov obtained the energy spectrum $E_k$ in the 
steady state of a fully grown turbulence as

$$E_k \sim k^{-\alpha}; ~~~ \alpha = 5/3, \eqno (1)$$
 
\noindent in dimension $d =$ 3, for an intermediate 
well-spread ``inertial range" of the wave vector $k$ [1].

This dispersion exponent $\alpha$ has been identified as the inverse Flory 
exponent $\nu$ for  linear polymers or of SAWs [2].
In three dimensions, if the vortex lines are assumed not to cross (as in
the absence of viscosity), the vortex line conformations can be modelled 
by those for SAWs. Hence the spatial distribution of  energy density $E(r)$
can be obtained from the SAW pair correlation $g(r)$  (at a distance $r$), 
and the Fourier
transform will then give 

$$E_k \sim g_k. \eqno (2)$$

\noindent 
The Fourier transform 
$g_k = g(ka,N)$ can be taken to be a function of the number of steps $N$ 
of the SAW (each step size being $a$) and of the dimensionless number $ka$. 
Since the SAW end-to-end distance average $R_N$ grows as  $aN^{\nu}$, 
where $\nu$ denotes the size exponent, $R_N$ remains invariant under the
scale transformation $a \to al^{\nu}$ and $N \to N/l$. Since the pair
correlation comes from the scatterings of $N-1$ other steps or monomers, it
scales with $N$ and hence

$$g(ka,N) = lg(kal^{\nu}, N/l). \eqno (3)$$

\noindent
Choosing $N/l  = 1$, one gets
$g_k = N\tilde {g}(kaN^{\nu}).$
\noindent Assuming now that $\tilde{g}(x) \sim x^{-\alpha}$, one gets

$$ g_k \sim k^{-\alpha} N^{1 -\nu\alpha} \sim k^{-\alpha}, ~~ {\rm if} 
~~ \alpha = 1/\nu. \eqno (4)$$

\noindent This then, together with eqn. (2), gives
eqn. (1) with $\alpha = 1/\nu$.

It is well known that the Flory estimate [3] for the polymer
size exponent $\nu$ is

$$\nu = 3/(2+d). \eqno (5)$$

\noindent
This comes from the minimization of the free energy $f(R)$ of a chain of
linear size (end-to-end distance)  $R$: $f = f_e + f_r$ where $f_e \sim
R^2/N$ denotes the Gaussian chain estimate of the elastic part and $f_r \sim
R^d c^2 \sim N^2/R^d$ denotes monomer-monomer repulsive part of the free 
energy ($c = N/R^d$ denoting the monomer concentration).
The minimisation of this $f$ with respect to $R$ gives $R\sim N^{\nu}$,
where this Flory estimate of the $\nu$ is given by (5).

In $d = 3$, one gets $\nu = 3/5$ and hence $\alpha = 5/3$ as Kolmogorov
obtained [2]. 
 In $d = 2$,
such a mapping (of the vortex lines to SAWs) does not exist and hence the
correspondence between the Kolmogorov exponent $\alpha$ and the polymer/SAW
size exponent $\nu$ (which of course is meaningful for the SAWs)
can not be checked. At $d= 4$, of course,
turbulence has not been
studied. 

However, checking in different (intermediate) dimensions could perhaps be
done.
In view of the wide ranging studies [4] of  turbulence in porous media,
 in connection with oil explorations etc, one can easily extend the above 
correspondence (between the Kolmogorov energy dispersion exponent $\alpha$
 and the polymer size exponent $\nu$) in porous media, which has been modelled
 very extensively and successfully by percolating fractals [5]. In fact, the 
conformational properties  of the linear polymers or SAWs in porous media 
have been extensively studied recently  and often the rigorous theoretical 
and experimental or extensive computational results do not 
comfortably match each other [6]. 
May be, the studies on the energy dispersion for turbulence in porous media,
the corresponding Kolmogorov exponent in particular, can also help in major
reconciliation for the polymer conformation studies in porous media.  

Since the fractal dimension $d_F$ of the porous rock is certainly less than 
that ($d$) of the embedding dimension ($d_F \simeq 2.5$ for percolation 
clusters in $d = 3$ [5]), following eqn. (5), 
 the Flory estimate $\nu_F$ is clearly
higher  than that ($\nu$ = 3/5) in the  normal media ($d =3$).
This immediately indicates that the Kolmogorov energy dispersion exponent 
$\alpha$ for turbulence in a porous medium will be considerably
below its standard value of 5/3. All the theoretical and computer simulation
results indicate convincingly [6] that the SAW size exponent $\nu_F$ on
 the percolation like fractals are larger than that ($\nu$)
 on the corresponding
Euclidean lattices (in the same embedding dimensions). This would clearly 
indicate a lower (than 5/3) Kolmogorov exponent ($\alpha$)
 value for turbulence in ($d=3$)
porous media. A recent approximate renormalization group study [7] 
of turbulence in porous media, however, seems to suggest unchanged 
value for the 
Kolmogorov exponent $\alpha$. In view of this, rigorous theoretical 
as well as experimental studies on turbulence in porous media
are needed to settle the issue.

\bigskip

\noindent {\bf Acknowledgement:} I am grateful to A. Basu and S. M. 
Bhattacharjee for useful discussions and criticisms, and to A. A. Avramenko 
for sending a copy of their paper.

\vskip 1 cm

\leftline {\bf References:}

\medskip 

\noindent [1] see e.g., C. C. Lin and W. H. Reid, {\it Turbulent Flow: 
Theoretical Aspects}, in {\it Handbuch der Physik}, Vol. II, S. Flagge
and C. Truesdele (Eds.), Springer, Berlin (1963).

\medskip

\noindent [2] see e.g., K. Huang, {\it Lectures on Statistical Physics and 
Protein Folding}, World Scientific, Singapore (2005).

\medskip

\noindent [3] see e.g., P. G. de Gennes, {\it Scaling Concepts 
in Polymer Physics},
Cornell Univ. Press, Ithaca (1979).

\medskip

\noindent [4] see e.g., J. L. Lage, J. M. S. de Lemos and D. A. Nield,
 {\it Modelling Turbulence in Porous Media}, in {\it Transport Phenomena in
Porous Media}, Vol. II,  D. B. Inghan and I. Pop (Eds.), Elsevier, 
Amsterdam (2002).

\medskip

\noindent [5] see e.g., D. Stauffer and A. Aharony, {\it Introduction 
to Percolation 
Theory}, Taylor \& Francis, London (1992).

\medskip

\noindent [6] see e.g., B. K. Chakrabarti (Ed.), {\it Statistics of Linear
Polymers in  Disordered Media}, Elsevier, Amsterdam (2005); K. Barat and
B. K. Chakrabarti, Phys. Rep. {\bf 258} 377-414 (1995).

\medskip 

\noindent [7] A. A. Avramenko and A. V. Kuznetsov, 
Transport in Porous  Media (Springer), {\bf 63} 175-193 (2006).

\end{document}